\documentclass[12pt]{iopart}
\begin{document}
\include{epsf}

\jl{3}      

\title{Droplet shapes on structured substrates and conformal invariance}

\author{A.O.\ Parry, E.D.\ Macdonald and C.\ Rasc\'{o}n}
\address{Department of Mathematics,\\
Imperial College, London. SW7 2BZ}

\date{\today}
\maketitle
\begin{abstract}
We consider the finite-size scaling of equilibrium droplet shapes for fluid adsorption (at bulk two-phase co-existence) on heterogeneous 
substrates and also in wedge geometries in which only a finite domain $\Lambda_{A}$ of the substrate is completely wet. For 
three-dimensional systems with short-ranged forces we use renormalization group ideas to establish that both the shape of the droplet 
height and the height-height correlations can be understood from the conformal invariance of an appropriate operator. This allows us to 
predict the explicit scaling form of the droplet height for a number of different domain shapes. For systems with long-ranged forces 
conformal invariance is not obeyed but the droplet shape is still shown to exhibit strong scaling behaviour. We argue that droplet 
formation in heterogeneous wedge geometries also shows a number of different scaling regimes depending on the range of the forces. 
The conformal invariance of the wedge droplet shape for short-ranged forces is shown explicitly.
\end{abstract}

\section{Introduction}
There is currently much experimental and theoretical interest in the
behaviour of fluid adsorption on structured substrates that are chemically
heterogeneous [1-5] or that have a specific non-planar shape [6-10]. It is now 
appreciated that breaking the translational invariance along the substrate can 
lead to a number of highly interesting effects. These include surface phases that
do not obey Youngs contact angle equation [1-3], surface condensation-like phase
transitions [4-6], non-trivial responses of the interface shape to substrate
corrugation \cite{7}, geometry dependent critical exponents for the growth of the
adsorption at complete wetting \cite{8} and even changes to the roughness exponent 
describing the fluctuations of the interface at filling transitions compared to
wetting \cite{9,10}. In the present work we are concerned with the equilibrium shape
of finite-size droplets on  planar substrates and also in
non-planar wedge geometries. We argue that within the grand canonical ensemble 
(without volume constraint) the droplet shape is characterised by universal 
scaling-like functions analogous to universal order-parameter profiles 
associated with finite-size effects in critical magnets and (binary) fluids 
\cite{11,12}. In particular for systems with short-ranged forces we show that the 
universal droplet shape may be understood from the conformal invariance of an 
appropriately defined surface operator. Indeed for such systems
the conformal invariance may be proved explicitly using a straightforward
renormalization group analysis. In this way we are able to make predictions for
the explicit droplet shape on a number of pertinent heterogeneous substrates 
including those with striped and circular domains. We believe that our 
predictions can be tested in future numerical simulation studies of Ising models
and may be open to experimental verification.

Our article is arranged as follows: In \S 2 we discuss the case of droplet
formation on planar substrates in (bulk) three-dimensional ($d \! = \! 3$) systems (i.e. a
two-dimensional interface) concentrating on the case of systems with
short-ranged forces (for which they may well be experimental systems) where the
conformal invariance of the droplet shape and its correlations may be proved 
directly and used to understand the finite-size scaling behaviour. We finish this 
section with a discussion of long-ranged forces (in $d \! = \! 3$) where the droplet
shape is characterised by alternative scaling functions and also a brief 
treatment of droplet formation in bulk dimension $d \! = \! 2$ where the result may also
be related to the restrictions imposed by conformal invariance. In \S 3 we identify
the scaling functions describing the shape of droplets in heterogeneous wedge geometries
exactly at the filling transition temperature for systems with both 
short-ranged and long-ranged dispersion forces. Again we find that the
restrictions imposed by conformal invariance have relevance to droplet
formation for short-ranged forces.

\section{Droplets on planar substrates}
\begin{figure}[c]
\begin{center}
\leavevmode \hbox{%
\epsfxsize=0.75\textwidth
\epsffile{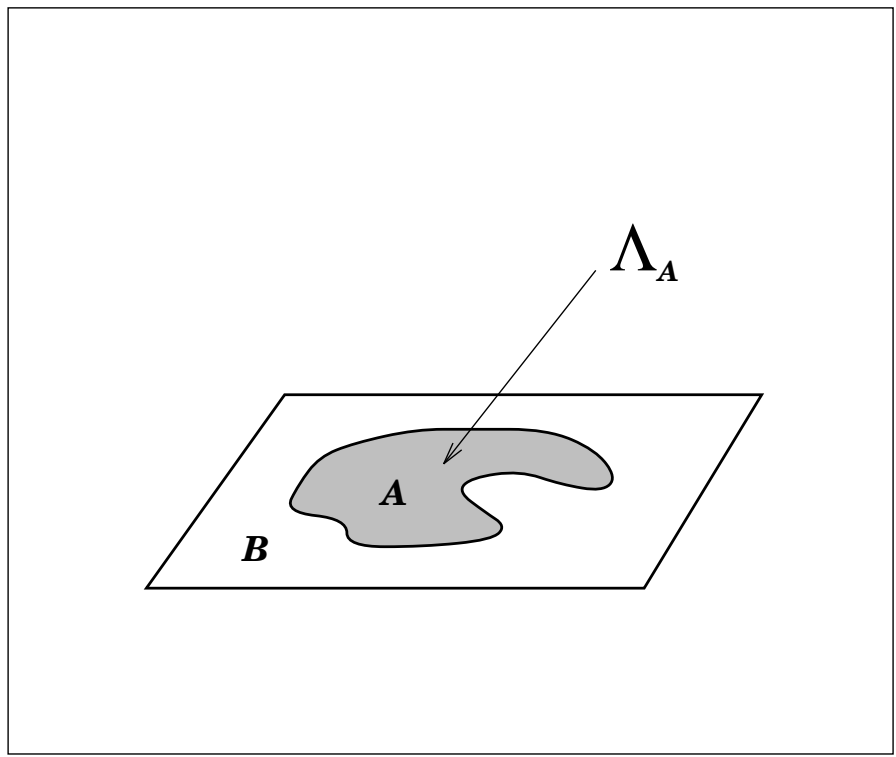}}
\end{center}
\caption{\label{Fig.1.} Schematic illustration of a heterogeneous planar substrate containing a finite-size domain of substrate type $A$ which is completely wet by liquid at bulk two-phase coexistence corresponding to local contact angle $ \theta_{A} \! = \! 0$. Outside this domain the substrate is of type $B$ and is only partially wet so $\theta_{B} \! > \! 0$. All lengthscales are considered larger than the bulk correlation length.}
\end{figure}

To begin we discuss the case of droplet shapes on planar heterogeneous
substrates. Consider a planar substrate in contact with a vapour phase
which is at bulk liquid-vapour coexistence corresponding to saturation
chemical potential $\mu \! = \! \mu_{sat}$. The substrate is chemically heterogeneous and consists
of a finite-size simply connected domain $\Lambda_{A}$ of substrate type $A$ (which is assumed 
to extend over a region much larger than the bulk correlation length in all 
directions) which would be completely wet by the liquid phase if the area was 
infinite i.e. the local contact angle $\theta_{A} \! =  \! 0$. Outside of the domain the substrate
is of type $B$ which is only partially wet by the liquid phase at this temperature
so that $\theta_{B} \! > \! 0$, see Fig.\ref{Fig.1.}. These conditions induce the condensation of a 
large finite-size droplet over the domain $\Lambda_{A}$ with the local height of the interface pinned to a 
finite value at the edges. An extreme example of this is if substrate $B$ has
contact angle $\theta_{B} \! = \! \pi$. This situation is realisable in experiments with water in 
which $\alpha$ is hydrophilic and $B$ is hydrophobic say. Alternatively for Ising-like 
magnets one may envisage fixing the surface spins lying within $\Lambda_{A}$  to be $+ 1$  and
the spins outside to $- 1$. As the area of the domain is increased so the 
equilibrium height of the droplet grows and becomes macroscopic. We are 
interested in the shape of the droplet for large domain sizes as this 
asymptotic limit is approached. In this limit the details of the wetting
properties of $B$ will not matter provided they serve to pin the interface to
a finite height at the domain boundary. Similarly we do not have to make any
assumptions regarding the order of any wetting transition occurring for the   
substrate. The central question that we address here is how the shape of the 
domain $\Lambda_{A}$ bounding the drop affects the equilibrium 
droplet shape. Note that there is no volume constraint imposed on the amount of
liquid adsorbed on the domain so the equilibrium shape of the interface
necessarily has the same translational invariance as the underlying substrate 
binding potential (see \cite{1,2} for effects induced by volume constraint or
equivalently by over-saturating the bulk vapour phase). The simplest possible 
starting point for discussing this problem which includes the effect of 
interfacial fluctuations and microscopic interactions is the interface 
Hamiltonian model \cite{4,5}
\begin{equation} \label{1}
H[\ell] = \int {\bf{d x}} \left \{ \frac{\Sigma}{2} (\nabla \ell)^{2} + W(\ell;{\bf{x}}) \right \}
\end{equation}
where $\ell({\bf{x}})$ denotes the local interface height at position ${\bf{x}} \! = \! (x,y)$ along the
substrate, $\Sigma$ is the interfacial stiffness (tension) of the unbinding fluid
interface and $W(\ell;{\bf{x}})$ denotes a local position dependent binding potential. As the
position ${\bf{x}}$ changes from outside to inside the domain so the local potential
changes from being of type $W_{B}(\ell)$ to $W_{A}(\ell)$ as appropriate to the respective
homogeneous $B$ and $A$ substrates. The cross-over in $W(\ell;{\bf{x}})$ can be anticipated to
occur over a finite distance $\lambda_{A \, B}$ \cite{4,5} which itself depends on the range of the
forces in the system. For short-ranged forces we anticipate that $\lambda_{A \, B}$ is of order
the bulk correlation length. This model also assumes that the interfacial
gradient is small otherwise the first term in (\ref{1}) has to be replaced with the
usual drumhead expression for the interfacial area. We will only be interested
in the scaling properties of the droplet shape however which emerge when the
gradient is small. For the model (\ref{1}) it is straightforward to derive
an exact Ward identity for the equilibrium interface position. Following the
analysis of Mikheev and Weeks \cite{13} (see also \cite{14}) it is straightforward to show
that
\begin{equation} \label{2}
\Sigma \nabla^{2} \langle \ell \rangle = \left \langle \partial W(\ell;{\bf{x}}) / \partial \ell \right \rangle 
\end{equation}
which relates the local effective force to the local interface curvature. 
Here $\langle \cdot \rangle$ denotes the usual ensemble average. Thus
provided we restrict our attention to the scaling limit of domain area $A_{\Lambda} \! \rightarrow \! \infty$ and
positions ${\bf{x}}$ which are far from the domain edge then the shape of the droplet
is described by the solution of
\begin{equation} \label{3}
\Sigma \nabla^{2} \langle \ell \rangle = \langle W^{\prime}_{A}(\ell) \rangle; \hspace{0.5cm} {\bf{x}} \in  \Lambda_{A}
\end{equation}
which serves as the starting point for our analysis. This scaling limit is 
analogous to that encountered in studies of universal order-parameter profiles 
in confined critical systems \cite{11,12}. To proceed we use the very well developed
renormalization group (RG) theory of wetting [15-19] and replace the RHS of (\ref{3})
with the derivative of an effective potential which is constructed on the
basis of the linearised RG. In other words we rewrite (\ref{3}) as 
\begin{equation} \label{4}
\Sigma \nabla^{2} \langle \ell \rangle = W_{ef\!f}^{\prime}(\langle \ell \rangle); \hspace{0.5cm} {\bf{x}} \in  \Lambda_{A}
\end{equation}
where the effective potential is derived by coarse-graining the bare potential 
over the local roughness of the interfacial fluctuations. The net result (in
dimension $d \! = \! 3$) is [15-19] 
\begin{equation} \label{5}
W_{ef\!f} ( \ell ) = \int_{- \infty}^{\infty} dt \, W(t) \frac{e^{- (\ell - t)^{2}/2 \xi_{\bot}^{2}}}{\sqrt{2 \pi \xi_{\bot}^{2}}}
\end{equation}
where $\xi_{\bot} \! = \! \sqrt{\frac{k_{B} T}{2 \Sigma \pi} \ln \xi_{\parallel}}$ is the local effective roughness and 
$\xi_{\parallel}$ is a transverse correlation length determined self-consistently 
by $\xi_{\parallel}^{-2} \! \propto \! W^{\prime \prime}_{ef\!f}(\ell)$. This then is the desired
field equation for the mean droplet height, the precise form of which
depends on the range of the forces. Note that the existence of the effective
potential implies that there exists a local free-energy functional $F[\ell]$ for the 
mean interfacial height (and its correlations) which is of the same form as (\ref{1})
(at least within the desired region ${\bf{x}} \! \in \! \Lambda_{A}$) but with $W_{A}(\ell)$ replaced with 
$W_{ef\!f}(\ell)$. The existence of a local free-energy functional for the droplet problem is directly
related to the fact that the exponent analogous to the bulk correlation 
function exponent $\eta$ is zero for wetting transitions. This has important 
simplifying consequences for the structure of correlations in the droplet which we
will discuss later.
 
To continue we consider the cases of short-ranged and long-ranged forces separately.

\subsection{Droplet shapes for short-ranged forces}
\begin{figure}[c]
\begin{center}
\leavevmode \hbox{%
\epsfxsize=0.85\textwidth
\epsffile{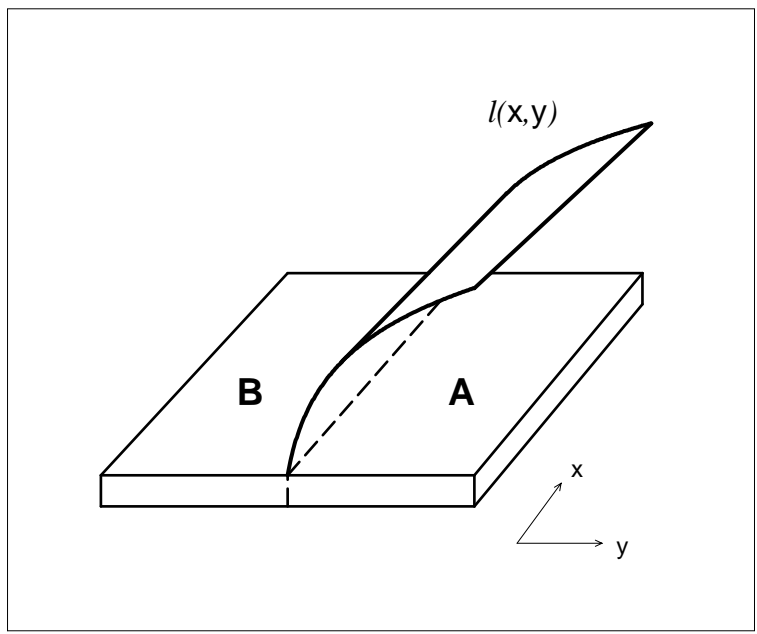}}
\end{center}
\caption{\label{Fig.2.} Heterogenous substrate made from the union of two semi-infinite surfaces of type $A$ and $B$. The mean interface height grows to infinity as we move further into the completely wet domain. For short ranged forces the function $e^{- \tilde{\kappa} \langle \ell \rangle}$ has the scaling form $e^{- \tilde{\kappa} \langle \ell \rangle} \! \propto \! y^{- 2}$ far from the $y \! = \! 0$ line.}
\end{figure}

\begin{figure}[c]
\begin{center}
\leavevmode \hbox{%
\epsfxsize=0.85\textwidth
\epsffile{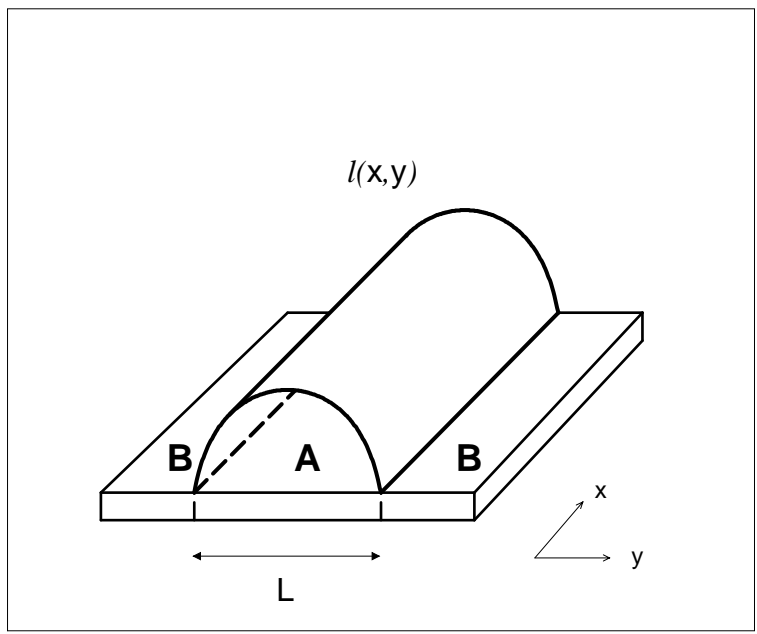}}
\end{center}
\caption{\label{Fig.3.} Schematic illustration of a droplet cross-section on a heterogenous striped domain of finite-width $L$. For short-ranged forces the analytic expression for the scaling of the droplet shape is obtained by conformally mapping the semi-infinite substrate shown in Fig.\ref{Fig.2.}.}
\end{figure}

For systems with short-ranged forces such as Ising and Landau-Ginzburg-Wilson
models and also some experimentally accessible systems including polymers \cite{20}, 
some fluid mixtures \cite{21} and superconductors \cite{22} it is well understood [15-19]
that the bare (unrenormalised) potential decays as $W_{A}(\ell) \! \sim \! A_{s} e^{- \kappa \ell}$ with $A_{s}$ an effective (positive) 
Hamaker constant and $\kappa \! \equiv \! 1 / \xi_{b}$ is the inverse bulk correlation length \cite{23,24}. We emphasise here that because 
the droplet shape reflects finite-size effects occurring at complete wetting we only need account for the
leading order term in the decay of the binding potential. The renormalization of this is very well understood and gives rise to an 
effective potential which depends on the value of the wetting parameter $\omega \! = \! \frac{k_{B} T \kappa^{2}}{4 \pi \Sigma}$. For 
values of $\omega \! < \! 2$ as is pertinent to fluids and Ising systems \cite{23,24} there is no need to consider the "hard wall" 
restriction $l \! > \! 0$ (which can only be treated approximately by the linear RG method) and the effective potential is $A_{s} e^{- \tilde{\kappa} \ell}$ where 
$\tilde{\kappa} \! = \! \kappa/(1 + \omega/2)$ is a renormalised inverse lengthscale. The field equation for three-dimensional 
droplets with short-ranged forces is therefore
\begin{equation} \label{6}
\Sigma \nabla^{2} \langle \ell \rangle = - A_{s} \tilde{\kappa} e^{- \tilde{\kappa} \langle \ell \rangle}
\end{equation}
which we wish to solve for different shapes of the domain $\Lambda_{A}$. The central
observation of this paper is that the above equation is conformally invariant
with $e^{- \tilde{\kappa} \langle \ell \rangle}$ playing the role of a primary operator with scaling dimension $x_{b} \! = \! 2$ \cite{25}. 
The proof of this is straightforward, but tedious and is presented in the appendix. Specifically under a conformal map 
$z \! \rightarrow \! w$, where $w \! = \! u + i v$ is an arbitrary analytic function of the complex variable 
$z \! = \! x + i y$, the field equation (\ref{6}) is invariant provided we identify
\begin{equation} \label{7}
e^{- \tilde{\kappa} \langle \ell \rangle (u, v)} = \vert w^{\prime} (z) \vert^{-2} e^{- \tilde{\kappa} \langle \ell \rangle (x, y)}.
\end{equation}
To exploit this result consider the simplest example of a structured substrate
consisting of two semi-infinite planes of type $A$, $B$ which meet at the line $y \! = \! 0$
(See Fig.\ref{Fig.2.}). From (\ref{6}) it follows that $e^{- \tilde{\kappa} \langle \ell \rangle}$ shows the scaling behaviour
\begin{equation} \label{8}
e^{- \tilde{\kappa} \langle \ell \rangle_{\infty/2}} = C y^{- 2}; \hspace{0.5cm}\tilde{\kappa} y \gg 1
\end{equation}
so that $\tilde{\kappa} \langle \ell \rangle_{\infty/2} \! \sim \! 2 \ln (y C^{-1/2})$. Here 
$C \! = \! 2 \Sigma / A \tilde{\kappa}^{2}$ is a non-universal constant. Following the analysis of 
Burkhardt and Eisenreigler \cite{26} we now use some well known analytic functions to conformally map the
semi-infinite system onto a number of pertinent finite-size domains and use (\ref{7})
to identify the appropriate droplet shape. For example the analytic functions $w \! = \! \frac{L}{\pi} \ln z$ and 
$w \! = \! \frac{L}{\pi} \cosh^{-1} z$ map the half plane onto the infinitely long strip 
$\{- \infty \! < \! u \! < \! \infty\} \cup \{0 \! < \! v \! < \! L\}$ and half-strip $\{u \! > \! 0 \} \cup \{0 \! < \! v \! < \! L\}$ 
respectively. In this way we obtain the droplet shapes as
\begin{equation} \label{9}
e^{- \tilde{\kappa} \langle \ell \rangle_{strip}} = C \left( \frac{L}{\pi} \sin \frac{\pi y}{L} \right)^{-2}
\end{equation}
for $0 < y < L$ and
\begin{equation} \label{10}
e^{- \tilde{\kappa} \langle \ell \rangle_{half strip}} = C \left[ \left( \frac{L}{\pi} \sinh \frac{\pi x}{L}\right)^{-2} + \left( \frac{L}{\pi} \sin \frac{\pi y}{L}\right)^{-2} \right]
\end{equation}
for $0 < y < L$ and $x > 0$. Note that in writing these expressions we have changed our co-ordinate notation back to $(x,y)$ rather than
persisting with $(u,v)$ (see Fig.\ref{Fig.3.}).
  
For circular domains of radius $R$ the special conformal group can be used to obtain the desired droplet shape 
\begin{equation} \label{11}
e^{- \tilde{\kappa} \langle \ell \rangle_{disc}} = C \left\{ \frac{R}{2} \left[1 - \left(\frac{r}{R}\right)^{2}\right] \right\}^{-2}
\end{equation}
where $r$ denotes the distance from the centre of the disc. Note that the special
conformal group can be used to map hyperplanes onto hyperspheres in arbitrary
dimension so that we may anticipate that a result analogous to (\ref{11}) may exist in
other dimensions in particular if the substrate (or unbinding interface) is 
effectively one dimensional. We shall return to this point later.

Equations (\ref{9}) and (\ref{11}) are our main predictions for droplet shapes in
systems with short-ranged forces. We believe these to be the most relevant for future
numerical and simulation studies and we finish this section with 
a discussion of them. Our first remark is that the droplet-shape shows the same 
kind of finite-size scaling behaviour as predicted for order-parameter profiles
in confined critical systems \cite{11,12}. Consider for example the result (\ref{9}) for  
the strip domain. For large (but finite) distances from the droplet edge at the
$y \! = \! 0$ line, the local height shows the "distant wall correction"
\begin{equation} \label{12}
\tilde{\kappa} \langle \ell \rangle_{strip} = \tilde{\kappa} \langle \ell \rangle_{\infty/2} - \frac{\pi^{2}}{3} \left(\frac{y}{L}\right)^{2} + \dots
\end{equation}
for $y/L \! \ll \! 1$. The power of $L$ here reflects the two-dimensional character of the
interface \cite{11}. Secondly our results (\ref{9}) and (\ref{11}) for the droplet shape can be 
conveniently recast in scaling form by re-writing them in terms of the maximum
height $\ell^{(m)}$ of the droplet occurring at the middle of the strip, disc domains. Thus for the strip we predict
\begin{equation} \label{13}
\kappa \left( \langle \ell \rangle_{strip} - \ell_{strip}^{(m)} \right) = (2 + \omega) \ln \sin \frac{\pi y}{L}
\end{equation}
whilst for the disc
\begin{equation} \label{14}
\kappa \left( \langle \ell \rangle_{disc} - \ell_{disc}^{(m)} \right) = (2 + \omega) \ln \left( 1 - \frac{r^{2}}{R^{2}} \right).
\end{equation}
Again we emphasise that the scaling function (\ref{13}) is valid for $y \rightarrow \infty$, 
$L \rightarrow \infty$ with $y/L$ arbitrary (and similarly for (\ref{14})).
It is these expressions that we believe would be easiest to test in future
simulation or experimental studies. Before we discuss the possibility of this we
mention here that in contrast to studies of order-parameter profiles in confined
critical magnets and fluids where the absolute value of the singular
contribution to the mid-point magnetisation (say) is very difficult to measure
(because it vanishes as $L \! \rightarrow \! \infty$ or $R \! \rightarrow \! \infty$ ) the scaling of the droplet height is
an obvious experimental observable. In particular the above results make a
precise prediction for the difference in the maximum (mid-point) droplet height
for the strip and disc domains. Specifically the difference in height of a droplet on an 
infinite strip of width $L$ and that on a disc of identical diameter satisfies
\begin{equation} \label{15}
\kappa \left( \ell_{strip}^{(m)} - \ell_{disc}^{(m)} \right) = (2 + \omega) \ln \frac{4}{\pi}
\end{equation}
which depends only on the value of the wetting parameter. It is remarkable how
small this height difference is. In mean-field approximation which ignores
fluctuation effects and has an effective value of $\omega \! = \! 0$ the numerical value of
the RHS of (\ref{15}) is only $0.4831$, i.e. the droplet heights are different by less 
than half a bulk correlation length. Close to the bulk critical point where $\omega$  
approaches a universal value $\omega_{c} \! \simeq \! 0.77$ \cite{23} this number increases to $0.67$ but
again the difference between the two heights is nearly negligible.

One of the first tests of these predictions would be to numerically study a
mean-field Landau-like model of adsorption on heterogeneous walls. The influence
of fluctuations on the droplet shape is rather small and the above results
are valid even in the mean-field limit of $\omega \! = \! 0$. Beyond mean-field level it would
also be straightforward to consider Ising model simulation studies similar to
those performed for wetting at homogeneous walls \cite{27,28}. We note that for this
case it may well be that the appropriate value of $\tilde{\kappa}$ is altered by the 
coupling of interfacial fluctuations \cite{29,30}. This is certainly needed to 
quantitatively explain simulation studies \cite{28} of (other) finite-size effects 
at complete wetting \cite{31}. One obvious experimental constraint on these
predictions is the role of gravity which both thins the height and dampens the
the fluctuations of the liquid-vapour interface. However this effect will only
become important if the strip width (say) is larger than the capillary length
for gravity damped interfacial modes. Using the value of the surface tension
of water at room temperature this implies that gravity may be neglected for 
strip widths or disc radii of less than a few micrometers.

\subsection{Height fluctuations and correlations in the droplet}
Next we turn our attention to the scaling properties of the height-height 
correlation function $S({\bf{x_{1}}}, {\bf{x_{2}}}) \! = \! \langle \ell({\bf{x_{1}}}) \, \ell({\bf{x_{2}}}) \rangle - \langle \ell({\bf{x_{1}}})\rangle \, \langle \ell({\bf{x_{2}}}) \rangle$. There are a number of ways of 
calculating this quantity but following the treatment above we use the effective
potential (\ref{5}) to derive a renormalised Ornstein-Zernike (OZ) equation which 
for short-range forces reads
\begin{equation} \label{16}
\left( - \Sigma \nabla^{2}_{\!\!x_{1}} + A_{s} \tilde{\kappa}^{2} e^{- \tilde{\kappa} \ell}\right) S({\bf{x_{1}}}, {\bf{x_{2}}}) = \delta ({\bf{x_{2}}} - {\bf{x_{1}}}).
\end{equation}
Here we point out the invariant properties of this equation as well its explicit
solution for the semi-infinite and strip geometries. On making use of the
covariant relation (\ref{7}) it is straightforward to show that this equation is 
invariant under a conformal map $z \! \rightarrow \! w(z)$ with the correlation function 
transforming simply as
\begin{equation} \label{17a}
S({\bf{u_{1}}}, {\bf{u_{2}}}) = S({\bf{x_{1}}}, {\bf{x_{2}}})
\end{equation}
where we have adopted the notation ${\bf{u}_{1}} \! = \! (u_{1}, v_{1})$. The absence of an explicit scaling
operator term in this result reflects the marginal logarithmic growth of the
droplet height away from the boundary for systems with short-ranged
heterogeneous forces. The transformation law for the height correlations is equivalent to the
statement that the (connected) two-point correlator of the primary operator $e^{- \kappa \ell}$ transforms as 
\begin{equation} \label{17b}
\langle e^{- \kappa \ell ({\bf{u}_{1}})}e^{- \kappa \ell ({\bf{u}_{2}})}\rangle_{G^{\prime}} = \vert w^{\prime} (z_{1}) \vert^{-2} \vert w^{\prime} (z_{2}) \vert^{-2} \langle e^{- \kappa \ell ({\bf{x}_{1}})}e^{- \kappa \ell ({\bf{x}_{2}})}\rangle_{G}
\end{equation}
where the subscripts $G^{\prime}$ and $G$ refer to ensemble averages with respect to the
different geometries under the conformal transformation. That (\ref{17a}) and (\ref{17b})
are equivalent again owes itself to the existence of a renormalised local
binding potential which accounts for fluctuation effects. Having established
that the height-height correlation function
transforms conformally it immediately follows from the analysis of Cardy \cite{32}
that on the semi-infinite planar  surface (Fig.\ref{Fig.2.}) the two-point function 
depends only on a single variable such that
\begin{equation} \label{18}
S_{\infty/2}({\bf{x_{1}}}, {\bf{x_{2}}}) = \Phi \left( \frac{(x_{1} - x_{2})^{2} + y_{1}^{2} + y_{2}^{2}}{2 y_{1} y_{2}} \right)
\end{equation}
The scaling function is then determined by substituting this result into the
OZ equation (\ref{16}) with $e^{- \tilde{\kappa} \ell}$ given by (\ref{7}). We find that $\Phi(t)$ satisfies
\begin{equation} \label{19}
(t^{2} - 1) \, \Phi^{\prime \prime} + 2 t \, \Phi^{\prime} - 2 \, \Phi = 0
\end{equation}
from which we can easily determine the important asymptotic behaviour. At short
distances ${\bf{x_{1}}} \! \rightarrow \! {\bf{x_{2}}}$ corresponding to  $t \! \rightarrow \! 1$ it follows that 
$\Phi \! \sim \! \ln (t - 1)$ implying that the local roughness $\xi_{\bot} \! \sim \! \sqrt{\ln y}$ precisely as expected for 
interfacial fluctuations at the marginal dimension $d \! = \! 3$. On the other hand in the limit of large transverse
separations $\vert {\bf{x_{2}}} - {\bf{x_{1}}} \vert \! \rightarrow \! \infty$ corresponding to $t \! \rightarrow \! \infty$ it 
follows form (\ref{13}) that $\Phi \! \sim \! t^{-2}$ which identifies the surface scaling exponent $\eta_{\parallel} \! = \! 4$ 
\cite{32,33}. It is this limit of the scaling function which is required in order to understand the asymptotic decay of correlations in 
the infinite finite width strip geometry. Using the mapping discussed above it immediately follows from (\ref{12}) that the 
height-height correlation function decays as
\begin{equation} \label{20}
S_{strip}({\bf{x_{1}}}, {\bf{x_{2}}}) \simeq \sin^{2} \left ( \frac{\pi y_{1}}{L} \right ) \sin^{2} \left ( \frac{\pi y_{2}}{L} \right ) e^{- 2 \pi \vert x_{2} - x_{1} \vert /L}
\end{equation}
for $\vert x_{2} - x_{1} \vert \! \gg \! L$. In writing this expression we have again chosen to use the
original co-ordinate notation rather than $(u, v)$. The above expression 
identifies the universal finite-size droplet correlation length $\xi_{\parallel} \! = \! L/2 \pi$. This identification is consistent with the amplitude-exponent relation $L/ \xi_{\parallel} \! = \! 2/ \pi \eta_{\parallel}$ \cite{11}.

We note here that the same result for $S({\bf{x_{1}}}, {\bf{x_{2}}})$ in the strip 
follows from explicit solution of the OZ equation (\ref{11}) using the derived 
expression for the droplet height (\ref{8}). This can be written as a spectral 
expansion
\begin{equation} \label{21}
S_{strip}({\bf{x_{1}}}, {\bf{x_{2}}}) = \sum_{n = 0}^{\infty} \psi_{n}^{*}(y_{1}) \psi_{n}(y_{2}) e^{- E_{n} \vert x_{2} - x_{1} \vert}
\end{equation}
with eigenvalues and eigenvectors determined from solution of 
\begin{equation} \label{22}
- \psi_{n}^{\prime \prime} + 2 \left ( \frac{L}{\pi} \sin \frac{\pi y}{L} \right )^{-2} \psi_{n} = E_{n}^{2} \psi_{n}
\end{equation}
subject to $\psi_{n}(0) \! = \! \psi_{n}(L) \! = \! 0$ and the orthonormality condition
\begin{equation} \label{23}
2 \Sigma \int_{0}^{L} dy \, \psi_{n}^{*}(y) \psi_{m}(y) = \delta_{n m} E_{n}^{-1}.
\end{equation}
The solution to (\ref{22}) is found using standard methods. For the eigenvalue
spectrum we find $E_{n} \! = \! (n + 2) \pi / L$ (with $n \! = \!0,\,1,\,2,\,\dots$) whilst the eigenvectors are         
\begin{equation} \label{24}
\psi_{n}(x) = N_{n} \sin^{2} \left( \frac{\pi x}{L} \right ) F \left [- \frac{n}{2}, \, 2 + \frac{n}{2}, \, \frac{1}{2}; \, \cos^{2} \frac{\pi x}{L} \right ]
\end{equation}
for $n = 0$, $2$, $4$, $\dots$ and
\begin{equation} \label{25}
\psi_{n}(x) = N_{n} \sin^{2} \left (\frac{\pi x}{L} \right ) \cos \frac{\pi x}{L} F \left [\frac{1}{2} - \frac{n}{2}, \, \frac{5}{2} + \frac{n}{2}, \, \frac{3}{2}; \, \cos^{2} \frac{\pi x}{L} \right ]
\end{equation}
for $n = 1$, $3$, $5$, $\dots$. Here $F[\alpha, \beta, \gamma; x]$ denotes the usual hypergeometric function. 
We note here that the fact that the
exact result for $S({\bf{x_{1}}}, {\bf{x_{1}}})$ in the strip geometry obtained by conformally mapping the
the semi-infinite correlation function also follows from the solution to a local
field equation (\ref{16}) is intimately related to the existence of a local effective
free-energy functional for complete wetting which in turn reflects the fact
that there is no exponent analogous to the $\eta$ occurring for bulk critical
systems. In studies of finite-size effects in strip (and other) geometries
occurring at bulk criticality (for which  $\eta \! \neq \! 0$) local field equations for the
pair correlation function derived from approximate local functional models
\cite{34} do not capture all the necessary scaling features required by conformal
invariance (although they are surprisingly accurate and highly reliable for
certain quantities \cite{35}).

As a final remark concerning height-height correlations in the strip geometry
we note that the existence of a renormalised local binding potential also
implies that the correlations in the strip satisfy the algebra of correlations
discussed by Parry and Swain \cite{36} in the somewhat different context of 
magnetisation functionals. Defining the Fourier transform of the correlation
function $\tilde{S}(y_{1}, y_{2}; {\bf{q}})$ by
\begin{equation} \label{26}
\tilde{S}(y_{1}, y_{2}; {\bf{q}}) = \int {\bf{d x_{21}}} e^{- i {\bf{q}} \cdot ({\bf{x_{2}}} - {\bf{x_{1}}})} S({\bf{x_{1}}},{\bf{x_{2}}})
\end{equation}
and in particular denoting the zeroth moment $\tilde{S}_{0}(y_{1}, y_{2}) \! = \! \tilde{S}(y_{1}, y_{2}; 0) $, it follows from the
local character of the renormalised effective free-energy functional that for 
any ordered points $0 \! \leq y_{1} \! \leq y_{2} \! \leq \! y_{3} \! \leq L$, the correlations satisfy
\begin{equation} \label{27}
\tilde{S}(y_{1}, y_{2}; {\bf{q}}) \, \tilde{S}(y_{2}, y_{3}; {\bf{q}}) = \tilde{S}(y_{1}, y_{3}; {\bf{q}}) \, \tilde{S}(y_{2}, y_{2}; {\bf{q}})\end{equation}
and
\begin{eqnarray} \label{28}
\left( \frac{\tilde{S}_{0}(y_{1}, y_{3})}{\ell^{\prime} (y_{1}) \ell^{\prime} (y_{3})}  -
\frac{\tilde{S}_{0}(y_{1}, y_{2})}{\ell^{\prime} (y_{1}) \ell^{\prime} (y_{2})} \right)
\left( \frac{\tilde{S}_{0}(y_{1}, y_{3})}{\ell^{\prime} (y_{1}) \ell^{\prime} (y_{3})}  -
\frac{\tilde{S}_{0}(y_{2}, y_{3})}{\ell^{\prime} (y_{2}) \ell^{\prime} (y_{3})} \right)
 = \\ 
\left( \frac{\tilde{S}_{0}(y_{1}, y_{1})}{\ell^{\prime} (y_{1}) \ell^{\prime} (y_{1})}  -
\frac{\tilde{S}_{0}(y_{1}, y_{2})}{\ell^{\prime} (y_{1}) \ell^{\prime} (y_{2})} \right)
\left( \frac{\tilde{S}_{0}(y_{3}, y_{3})}{\ell^{\prime} (y_{3}) \ell^{\prime} (y_{3})}  -
\frac{\tilde{S}_{0}(y_{2}, y_{3})}{\ell^{\prime} (y_{2}) \ell^{\prime} (y_{3})} \right) \nonumber
\end{eqnarray} 
where $\ell^{\prime} (y)$ denotes the derivative of the equilibrium height profile given by (\ref{9}). These relations severely restrict the possible form of $\tilde{S}_{0}(y_{1}, y_{2})$. In particular they imply that the dimensionless quantity
\begin{equation} \label{29}
\sigma(y_{1}, y_{2}) \equiv \frac{\tilde{S}_{0}(y_{1}, y_{2})}{\ell^{\prime} (y_{1}) \ell^{\prime} (y_{2})} \, \frac{1}{k_{B} T} \, \frac{\partial^{2} \Delta F}{\partial L^{2}}
\end{equation}
where $\Delta F(L)$ denotes the excess or finite-size contribution to the droplet  free-energy, satisfies
\begin{equation} \label{30}
\sigma(y_{1}, y_{2}) = \left (1 - \sqrt{1 - \sigma(y_{1}, y_{1})} \right ) \left(1 \pm \sqrt{1 - \sigma(y_{2}, y_{2})} \right); \hspace{0.5cm}y_{1} < \frac{L}{2}
\end{equation}
and the $\pm$ signs refer to whether $y_{1}$ and $y_{2}$ are on the same/opposite side of the
mid-line at $y \! = \! L/2$. We emphasise here that the results (\ref{27})-(\ref{30}) are valid for
any choice of effective binding potential (\ref{5}) and are not specific to the
exponential form for systems with short-ranged forces. The central 
prediction (\ref{30}) of the correlation function algebra relates the correlation 
function structure within the droplet to the excess free-energy $\Delta F(L)$. For the
droplet with short-ranged forces $\Delta F(L)$ is readily calculated using the
appropriate effective free-energy functional. We find
\begin{equation} \label{31}
\Delta F (L) = \frac{\pi}{8 \omega} \, (2 + \omega)^{2} \, L^{-1}.
\end{equation}
Again the width dependence of this quantity is precisely in accord with
dimensional (hyperscaling) ideas for confinement in critical systems \cite{12}. Note
that the (Casimir) amplitude here is strongly non-universal and depends on the value of $\omega$. This is in contrast to 
finite-size effects at bulk criticality (in Ising systems for example) where the amplitude of $\Delta F (L)$ is universal 
and related to the central charge \cite{12} characterising the bulk universality class. The derivative of the free-energy determines 
the force  of solvation between the edges of the strip which might be measurable in force-balance experiments if the width $L$ 
of the droplet can be controlled externally. 

\subsection{Droplets shapes and correlations with long-ranged forces}
In the last section we explicitly showed and exploited the conformal 
invariance of the droplet shape and its correlations for systems with 
short-ranged forces. Whilst some examples of such systems
may well be found in the laboratory most solid-fluid and fluid-fluid systems 
will de dominated by dispersion forces for which the bare binding potential
(of the completely wet $A$ region) decays like $W_{A} \! \sim \! A_{l} \ell^{- 2}$ where again $A_{l}$ is an
effective (positive) Hamaker constant. It is well understood
\cite{37} that this potential is not changed by fluctuations in $d \! = \! 3$ so that the
field equation for the droplet shape is simply
\begin{equation} \label{32}
\Sigma \nabla^{2} \langle \ell \rangle = - 2 A_{l} \langle \ell \rangle^{-3}.
\end{equation}
This equation is certainly not invariant under the same kind of conformal
mapping as valid for systems with short-ranged forces consistent with general
expectations of conditions for conformal invariance. In order to investigate the
possible scaling properties of different finite-size droplet shapes it is
therefore necessary to solve (\ref{32}) explicitly for different domains. Here we
concentrate on the infinite strip and disc domains only. For the strip geometry
the differential equation is trivially solved and shows that the equilibrium
height profile has a simple scaling form
\begin{equation} \label{33}
\langle \ell \rangle_{strip} = 2 \ell_{strip}^{(m)} \sqrt{\frac{y}{L} \left( 1 - \frac{y}{L} \right)}; \hspace{0.5cm} 0<y<L
\end{equation}
where $\ell_{strip}^{(m)} \! \propto \! L^{1/2}$ is the maximum (mid-) height of the drop. Thus whilst there
is no conformal invariance for the droplet with dispersion forces there is
still a strong degree of scaling in this system. Indeed the scaling of the
height profile for long ranged forces is stronger than that predicted for
short ranged forces as can be seen from comparison with (\ref{13}). Notice that near
the edges of the droplet the interface height behaves as $\ell_{strip} \! \sim \! y^{1/2}$ (for $y/L \! \rightarrow \! 0$ say) where
the power law here can be identified with the ratio of complete wetting 
exponents $\beta_{co}/ \nu_{\parallel}$ where $\beta_{co}$ is the adsorption critical exponent and $\nu_{\parallel}$ is the
transverse correlation length critical exponents. For systems with dispersion
forces recall that $\beta_{co} \! = \! 1/3$ and $\nu_{\parallel} \! = \! 2/3$ consistent with the
explicit shape dependence give above (see for example \cite{18}).

Correlations along the droplet show a similar scaling structure. From the OZ equation 
\begin{equation} \label{34}
\left[ - \Sigma \nabla_{\!\!{\bf{x_{1}}}}^{2} + W_{ef\!f}^{\prime \prime}(\langle \ell \rangle) \right] S({\bf{x_{1}}}, {\bf{x_{2}}}) = \delta({\bf{x_{2}}} - {\bf{x_{1}}})
\end{equation}
it follows from the spectral analysis given earlier that $S({\bf{x_{1}}}, {\bf{x_{2}}})$ decays asymptotically as
\begin{equation} \label{35}
S({\bf{x_{1}}}, {\bf{x_{2}}}) \sim \psi_{0}(y_{1}) \psi_{0}(y_{2}) e^{- E_{0} \vert x_{2} - x_{1} \vert}; \hspace{0.5cm} \frac{\vert x_{2} - x_{1} \vert}{L} \rightarrow \infty,
\end{equation}
where $\psi_{0}$ and $E_{0}$ are found from numerical solution of the eigenvalue 
problem
\begin{equation} \label{36}
- \psi_{0}^{\prime \prime} + \frac{3 L^{2}}{4 y^{2} (L^{2} - y^{2})} \, \psi_{0} = E_{0}^{2} \, \psi_{0}
\end{equation}
with $\psi_{0}(0) \! = \! \psi_{0}(L) \! = \! 0$. In this way we find that the finite-size correlation 
length characterising height fluctuations along the droplet satisfies
$L / \xi_{\parallel} \! \simeq \! 5.25$ which 
is universal in the sense that it does not depend on the value
of the surface tension or Hamaker constant. Interestingly the ratio is smaller 
than that derived earlier for short-ranged forces in keeping with general expectations 
about the influence of long-ranged forces on the range of correlations.
 
\begin{figure}[c]
\begin{center}
\leavevmode \hbox{%
\epsfxsize=0.85\textwidth
\epsffile{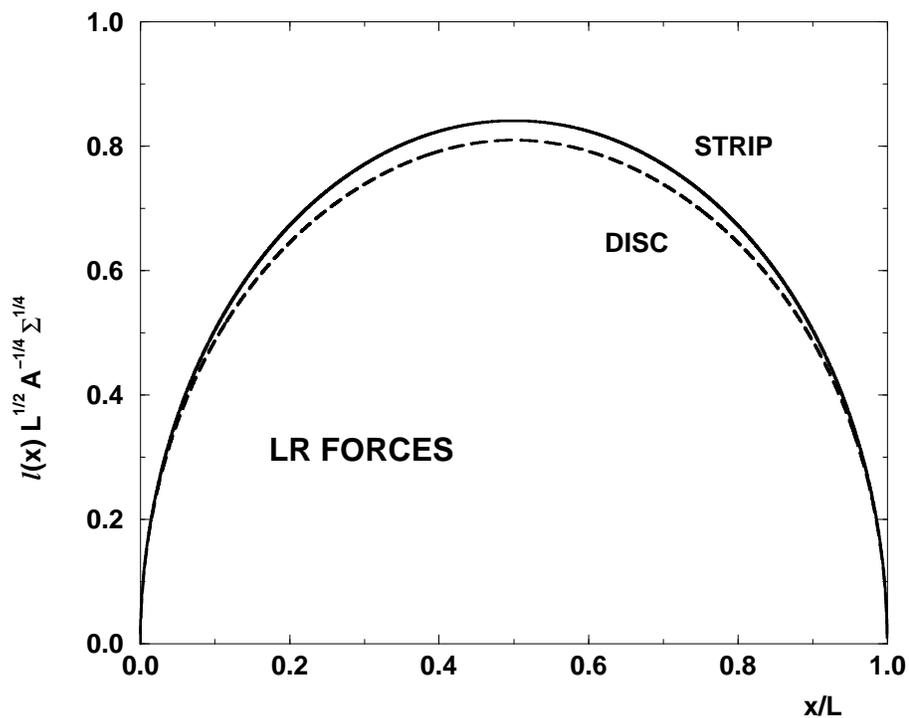}}
\end{center}
\caption{\label{Fig.4.} Comparison of the scaled interfacial height profiles for the strip and disc geometries for systems with dispesion forces. The disc radius $R$ is equal to the strip half-width. The height is measured in appropriate dimensionless units whilst the scaling variable $x/L$ denotes both the scaled distance across the strip and the radial distance across the disc.}
\end{figure}

For droplet shapes on circular domains of radius $R$ it follows from (\ref{32}) that
the equilibrium height profile also has a scaling form 
$\langle \ell \rangle_{disc} \! = \! \ell_{disc}^{(m)} \Lambda(r/R)$ 
where $\ell_{disc}^{(m)} \! \propto \! R^{1/2}$ is the mid-point height and $\Lambda(\xi)$ is a scaling function satisfying
 $\Lambda(0) \! = \! 1$ and $\Lambda(1) \! = \! 0$ which can be determined numerically. 
Following our earlier treatment of droplets with short-ranged forces it is
interesting to compare the maximum droplet height for a circular domain of
diameter $2 R \! = \! L$ with the corresponding droplet height for a strip
geometry. Because there is a stronger manifestation of scaling for systems with
long-ranged forces the comparison takes the form of a ratio 
$\ell_{strip}^{(m)} / \ell_{disc}^{(m)} \! \simeq \! 1.03$ 
rather than a difference as for short-ranged forces (\ref{15}). The influence of the
domain shape on the droplet adsorption is therefore stronger for systems
with dispersion forces even though the scaled heights are again very close to each-other. 
A comparison of the corresponding droplet shapes is shown in Fig.\ref{Fig.4.} for strips and discs with long-ranged forces.

\subsection{Droplet shapes in $1 + 1$ dimensions}
To end this section and motivate out treatment of droplet shapes in
three-dimensional wedges we remark here that the scaling form of the droplet
height for adsorption on a one-dimensional heterogeneous substrate also shows
scaling behaviour that can be understand using conformal invariance. To be
concrete consider a two dimensional semi-infinite Ising model at temperature $T$ 
(less than the bulk critical value $T_{c}$) and in zero magnetic field $H \! = \! 0^{-}$ (so that 
the bulk phase has negative magnetisation). The surface spins on the $y \! = \! 0$ line are
all fixed to be down except along the line between $x \! = \! 0$ and $+ L$ where they 
are all up. This is the two-dimensional analogue of the striped heterogeneous 
surface and the circular domain considered above. These boundary conditions 
induce the formation of a droplet of local equilibrium height $\ell(x)$ which is 
pinned to the wall at the end points $x \! = \! 0$ and $x \! = \! L$. In the limit of $L \! \rightarrow \! \infty$ the droplet
becomes macroscopic in size and we can anticipate that the shape shows scaling behaviour. The interfacial model for this problem is 
\begin{equation} \label{41}
H[\ell] = \int_{0}^{L} dx \left \{ \frac{\Sigma}{2} \left (\frac{d \ell}{d x} \right)^{2} + W(\ell) \right \}
\end{equation}
where the binding potential $W(\ell)$ may be simply approximated by an infinite
hard-wall potential $W(\ell) \! = \! 0$ for $\ell \! > \!0 $  and  $W(\ell) \! = \! \infty$ otherwise, which simply restricts the
interface to the $l \! > \! 0$ region. The reason for this is that in $d \! = \! 2$ the entropic
repulsion of the interface from the hard-wall dominates over any short-ranged
interaction (for complete wetting problems) [17-19]. This problem has been
considered by Burkhardt \cite{38} using continuum transfer-matrix techniques with the
scaling behaviour of $\ell(x)$ explicitly calculated to be
\begin{equation} \label{42}
\langle \ell(x) \rangle = 2 \ell^{(m)}_{0} \sqrt{\frac{x}{L} \left ( 1 - \frac{x}{L} \right)}
\end{equation}
where $\ell^{(m)}_{0} \! \propto \! L^{1/2}$ denotes the mid-point droplet height. Note that this
expression is identical to that derived above for $2 + 1$ dimensional systems with 
dispersion forces. Here we make two relevant remarks concerning this transfer 
matrix result :

(i) the scaling of $\langle \ell(x) \rangle$ in two dimensions is precisely what one would expect due
to the entropic repulsion from the hard-wall. The droplet shape reflects
interfacial finite-size effects in the weak-fluctuation regime (WFL) \cite{17,18}
which can be quantitatively understood using an effective potential which decays
as $W_{FL} \! \propto \! A_{FL} \ell^{-2}$. Using this form for the effective potential the field equation (\ref{4}) reduces to
\begin{equation} \label{43}
\Sigma \langle \ell \rangle^{\prime \prime} = - 2 A_{FL} \langle \ell \rangle^{-3}
\end{equation}
which is trivially integrated to recover the exact transfer matrix result (\ref{42}).

(ii) the scaling of $\langle \ell(x) \rangle$ is also in accord with the predictions of conformal
invariance for (non-trivial) one-dimensional field theories. In general phase
transitions do not occur in (classical) one-dimensional systems unless either
extremely long-ranged forces are present or the order-parameter is allowed to
diverge. The study of droplet shapes in $1 + 1$ dimensional systems is therefore an
interesting test of the ideas of conformal invariance in this dimension. To see
this first consider the limit of $L \! \rightarrow \! \infty$. Then far from the pinned edge of the droplet the
height $\ell(x)$ scales as
\begin{equation} \label{44}
\langle  \ell(x) \rangle \sim D x^{\zeta}; \hspace{0.5cm} \kappa x \gg 1
\end{equation}
where $\zeta \! = \! 1/2$ is the usual interfacial wandering exponent for the one-dimensional 
interface \cite{18,19}. This behaviour is consistent with the homogeneous rescaling $ x \! \rightarrow \! x/b$
under-which $\langle \ell (x) \rangle$ transforms as
\begin{equation} \label{45}
\left \langle  \ell \left (x/b \right ) \right \rangle = b^{- \zeta} \langle \ell(x) \rangle.
\end{equation}
It is natural to generalise this statement to non-uniform rescalings
corresponding to the one-dimensional limit of special conformal mappings 
such that
\begin{equation} \label{46}
\langle  \ell \rangle_{G^{\prime}} = \left ( \frac{d x^{\prime}}{d x}\right )^{\zeta} \langle  \ell(x) \rangle_{G},
\end{equation}
where as before $G^{\prime}$ and $G$ refer to ensemble averages with respect to the old and
new geometries respectively. The allowed class of transformations combines
uniform rescalings, translations and inversions is therefore of the form
\begin{equation} \label{47}
x^{\prime} = \frac{a + b x}{c + d x}
\end{equation}
with  $a d - b c \! \neq \! 0$ (since otherwise the scale factor vanishes). This transformation
maps the semi-infinite line to a line of finite length. Choosing $a$, $b$, $c$ and $d$ such that $0 \! < \! x^{\prime} \! < \! L$ we find
\begin{equation} \label{49}
\langle  \ell (x^{\prime})\rangle = \ell_{0}^{(m)} 4^{\zeta} \left [ \frac{x^{\prime}}{L}\left (1 -  \frac{x^{\prime}}{L} \right ) \right ]^{\zeta}
\end{equation}
which is of course identical to the transfer matrix result (\ref{42}) provided we
identify the correct value of the wandering exponent. Note that if we write this
expression in terms of the distance $x^{\prime \prime} \! = \! x^{\prime} - L/2$  from the centre of the strip then the
position dependence of the scaling function is basically identical to that
observed for circular domains in $d \! = \! 3$ but with a different value of $\zeta$ (see
(\ref{14})). This is because the scaling function for the height of radially symmetric 
drops on heterogeneous substrates (with short-ranged forces) in arbitrary 
dimension can be obtained by the application of the special conformal mapping analogous to (\ref{47}).
 
To finish this section we emphasise here that the only ingredient in deriving
the prediction (\ref{49}) is the assumption of co-variance under the special
conformal mapping (\ref{47}). The finite-size droplet in $1 + 1$ dimension is a specific
example of this one dimensional prediction of conformal invariance for a problem
in which $\zeta \! = \! 1/2$. In the next section we shall show that (\ref{49}) has application for a second example of interfacial phenomena. 

\section{Droplet shapes in three-dimensional wedge geometries}
The calculations presented in section \S 2 assume that the substrate is
heterogeneous but still planar. However recently it has been shown that
chemically homogeneous but non-planar substrates in the shape of wedge geometries
exhibit filling (or wedge-wetting transitions) which are characterised by very 
strong interfacial fluctuations belonging to quite distinct universality classes
and fluctuation regimes to those occurring for wetting at a planar wall \cite{10}.
The purpose of this section is to investigate the scaling of droplet shapes in 
wedge geometries. Before we study these finite-size effects we briefly review
the phenomenology of filling transitions and the recently developed fluctuation
theory for them based on analysis of a novel interfacial Hamiltonian \cite{10}. This
model will form the starting point for our study of droplet shapes in wedges.

\subsection{Filling transitions and effective Hamiltonian theory}
\begin{figure}[c]
\begin{center}
\leavevmode \hbox{%
\epsfxsize=0.85\textwidth
\epsffile{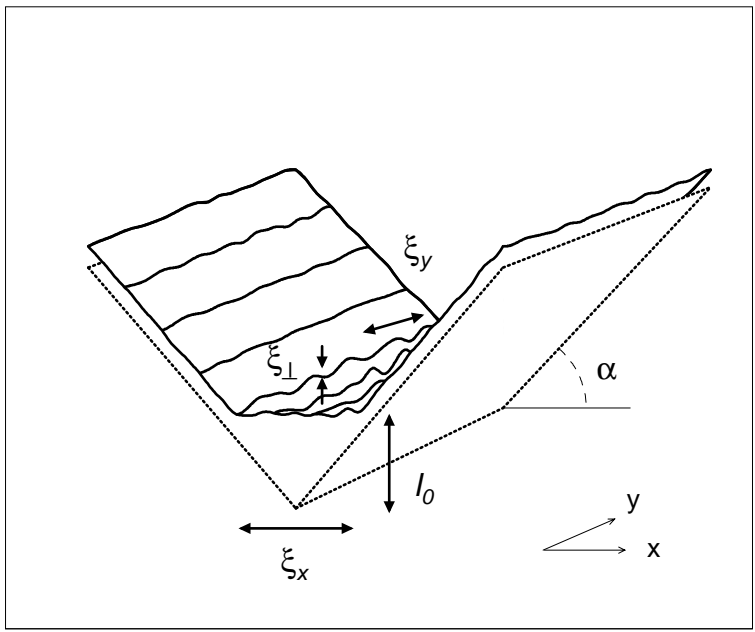}}
\end{center}
\caption{\label{Fig.5.} Section of an interfacial configuration occurring in a wedge geometry. As the critical filling transition temperature $T_{W}$ is approached the mean mid-point height $\ell_{0}^{max}$, roughness $\xi_{\bot}$ and anisotropic correlation lengths $\xi_{x}$, $\xi_{y}$ diverge continuously. The extreme anisotropy of fluctuations means that the transition is quasi-one dimensional.}
\end{figure}

\begin{figure}[c]
\begin{center}
\leavevmode \hbox{%
\epsfxsize=0.75\textwidth
\epsffile{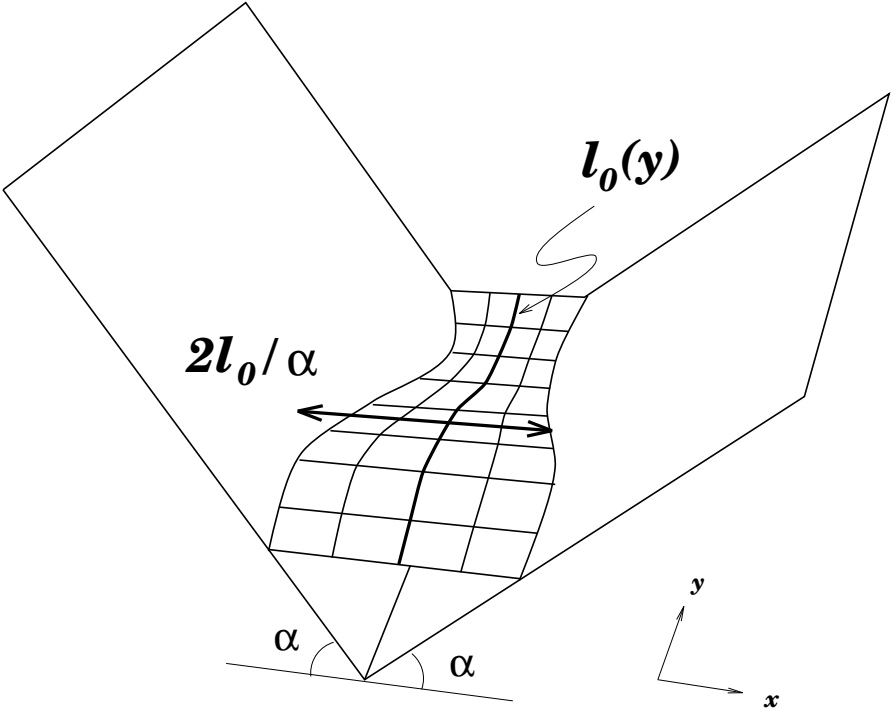}}
\end{center}
\caption{\label{Fig.6.} Schematic illustration of an interfacial configuration in the wedge geometry viewed on a scale larger than $\xi_{x}$. Fluctuations in the local height across the wedge have disappeared leaving only the quasi-one dimensional undulations of the mid-point height along the system.}
\end{figure}

Consider a wedge formed by the union of two walls at angles $\pm \alpha$ to the
horizontal. The wedge is supposed to be in contact with a bulk vapour phase at 
saturation chemical potential $\mu \! = \! \mu_{sat}(T)^{-}$ and is at a temperature $T$ less than the
wetting transition temperature $T_{w}$ of the planar wall-vapour
interface. Thermodynamic arguments [39-41] dictate that the wedge is completely
filled with liquid at a temperature $T_{F}$ satisfying     
\begin{equation} \label{50}
\theta(T_{F}) = \alpha
\end{equation}
and $\theta(T)$ denotes the contact angle of the liquid drop on the planar substrate. This
thermodynamic prediction is confirmed by specific model calculations which
include the influence of intermolecular interactions and fluctuation effects
\cite{9,10}. The filling transition occurring as $T \! \rightarrow \! T_{F}$ may 
be first or second-order \cite{42} 
corresponding to the discontinuous or continuous divergence of the interfacial 
height $\ell_{0}$ as measured from the bottom of the wedge. Importantly even substrates 
that exhibit first-order wetting transitions can exhibit continuous filling 
transitions \cite{10} so that the experimental observation of critical (continuous)
filling is a realistic possibility. In addition to the divergence of $\ell_{0}$           
as $T \! \rightarrow \! T_{F}^{-}$, the filling transition is characterised by the divergence of distinct
correlation lengths along $\xi_{y} \! \sim \! (T_{F} - T)^{- \nu_{y}}$ and across 
$\xi_{x} \! \sim \! (T_{F} - T)^{- \nu_{x}}$ the wedge, 
as well as the roughness $\xi_{\bot}$ of the unbinding interface (see Fig.\ref{Fig.5.}) \cite{10}. The mean-field critical
behaviour has been examined in detail \cite{10} from analysis of the non-planar
version of the standard interfacial Hamiltonian (valid for small $\alpha$) given by
\begin{equation} \label{51}
H[l] = \int {\bf{dx}} \left \{ \frac{\Sigma}{2} \left (\nabla \ell \right)^{2} + W(\ell - \alpha \vert x \vert) \right \}
\end{equation}
where $\ell({\bf{x}})$ is the local height of the interface as measured from the plane and $W(l)$ is the usual binding potential. 
The mean-field critical exponents for filling are determined only by the leading order decay term in $W(\ell)$. Writing
this as $W(\ell) \! = \! - A_{l} \ell^{- p}$ (so that $p \! = \! 2$ corresponds to non-retarded dispersion forces) the
critical exponents are given by
\begin{equation} \label{52}
\beta_{0} = \frac{1}{p}\hspace{0.25cm};\hspace{0.5cm}
\nu_{x} = \frac{1}{p}\hspace{0.25cm};\hspace{0.5cm}
\nu_{y} = \frac{1}{2} + \frac{1}{p}\hspace{0.25cm};\hspace{0.5cm}
\nu_{\bot} = \frac{1}{4}.
\end{equation}
Notice the remarkable universal value of the roughness critical exponent. 
Analysis of the height-height correlations at filling show that fluctuations 
across the wedge are highly localised to the filled region with the local height 
$\ell({\bf{x}})$ at position $y$ along the wedge fluctuating coherently, i.e. the effective 
stiffness across the filled portion of the wedge is infinite. In this sense
fluctuations across the wedge are trivial which can be seen in the explicit
values of the critical exponents since $\beta_{0} \! = \! \nu_{x}$  for all values of $p$. On the 
other hand correlations along the wedge are characterised by a soft mode and
completely dominate the fluctuation behaviour. On the basis of this Parry, Rasc\'{o}n
and Wood \cite{10} have argued that the fluctuation theory of three-dimensional 
filling transitions can be understood using the effective one-dimensional model
\begin{equation} \label{53}
H_{F}[\ell_{0}] = \int dy \left \{ \frac{\Sigma \ell_{0}}{\alpha} \left (\frac{d \ell_{0}}{d y} \right)^{2} + V(\ell_{0}) \right \}
\end{equation}
where $\ell_{0}(y) \! = \! \ell(0, y)$ is the local height of the interface at the centre of the
wedge. This model is based on the assumption that only fluctuations along the
wedge contribute to the asymptotic critical behaviour (see Fig.\ref{Fig.6.}). This effective model is only presumed correct for 
the soft-mode fluctuations in the $y$ direction which determine the asymptotic critical behaviour. The model 
is not valid for wavevectors $Q \! > \! Q_{max} \! \simeq \! 1/\xi_{x}$. At this scale the full 2$d$ model 
(\ref{51}) has to be used. Notice that the
coefficient of the bending term is proportional to the local height of the
filled region (quite unlike the stiffness term for planar fluid interfaces)
reflecting the fact that a local increase in the height also increases the 
width of the filled region in the $x$ direction. The direct filling potential
appearing in (\ref{53}) is found to be
\begin{equation} \label{54}
V(\ell_{0}) = \frac{\Sigma}{\alpha}(\theta^{2} - \alpha^{2}) + A_{F} \ell^{1 - p} + \dots
\end{equation}
the minimum of which identifies the mean-field $\ell_{0}$. This filling Hamiltonian can
be studied exactly using transfer matrix techniques yielding a complete
classification of the possible critical behaviour. For $p \! < \! 4$ mean-field theory is
valid, whilst for $p \! > \! 4$ the critical behaviour is universal with critical
exponents given by
\begin{equation} \label{55}
\beta_{0} = \frac{1}{4}\hspace{0.25cm};\hspace{0.5cm}
\nu_{x} = \frac{1}{4}\hspace{0.25cm};\hspace{0.5cm}
\nu_{y} = \frac{3}{4}\hspace{0.25cm};\hspace{0.5cm}
\nu_{\bot} = \frac{1}{4}.
\end{equation}
Notice that in the fluctuation dominated regime $\beta_{0} \! = \! \nu_{\bot}$ so there are regular
excursions of the interface to the wedge bottom. The existence of two 
fluctuation regimes for critical filling can be understood using arguments 
analogous to the well developed scaling theory of wetting [17-20]. To see this we
suppose that the influence of fluctuations at critical filling can be understood
using an effective potential
\begin{equation} \label{56}
V_{ef\!f}(\ell_{0}) = V(\ell_{0}) + V_{FL}(\ell_{0})
\end{equation}
which accounts for fluctuation effects by adding to the direct potential a
term arising from the entropic repulsion form the wedge bottom. The
fluctuation term can be estimated from the form of the bending energy
contribution to (\ref{53}) as $V_{FL}(\ell_{0}) \! \simeq \! \ell_{0}^{3} \xi_{y}^{- 2}$ and hence 
\begin{equation} \label{57}
V_{FL}(\ell_{0}) \simeq \ell_{0}^{3 - 2/\zeta}
\end{equation}
where $\zeta$ is the wandering exponent for filling transitions which relates the
roughness $\xi_{\bot} \! \sim \! \xi_{y}^{\zeta}$ and soft-mode correlation length $\xi_{y}$. The value $\zeta \! = \! 1/3$ can be 
either read directly from the transfer matrix results (\ref{55}) or equivalently
from the scale invariance of the free filling Hamiltonian $H_{0} \! = \! \int dy \frac{\Sigma \ell_{0}}{\alpha} \left (\frac{d \ell_{y}}{d y} \right)^{2}$ under the scale transformation $y \! \rightarrow \! y/b$ and $\ell \! \rightarrow \! \ell/b^{\zeta}$. The net result of this is an
effective filling potential of the form
\begin{equation} \label{58}
V_{eff}(\ell_{0}) = V(\ell_{0}) + D \ell_{0}^{- 3}
\end{equation}
which is minimised to find the equilibrium film thickness $\ell_{0}$. Thus we find a
mean-field and fluctuation dominated regime depending on whether the entropic
repulsion or direct algebraic term $A_{F} \ell^{1 - p}$ is the dominant correction to the
linear term. Unlike critical wetting
for which there are three fluctuation regimes critical filling only has two
since the linear term in the direct binding potential is always relevant. In
this sense fluctuation effects at three dimensional critical filling is rather
similar to those occurring at two-dimensional complete wetting [17-20,37].

\subsection{Finite-size droplets in heterogeneous wedge geometries}
\begin{figure}[c]
\begin{center}
\leavevmode \hbox{%
\epsfxsize=0.75\textwidth
\epsffile{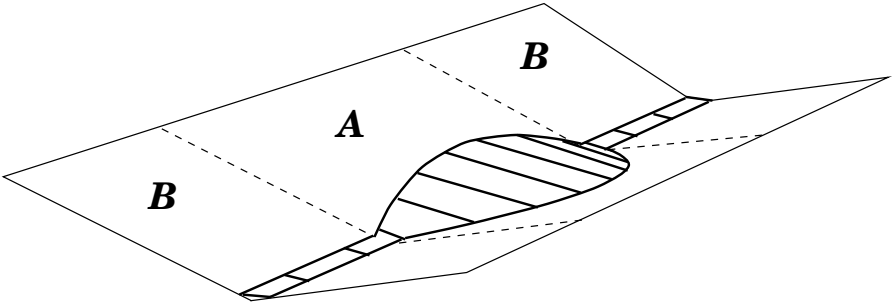}}
\end{center}
\caption{\label{Fig.7.} A heterogenous wedge containing a finite-size region $A$ which satisfies the local filling condition $\theta_{A}(T_{F}) \! = \! \alpha$. Outside this strip, in the region $B$ the wedge is only partially filled.}
\end{figure}

We are now in a position to consider the finite-size scaling of droplet 
shapes in heterogeneous wedge geometries formed by sandwiching a finite length
wedge of substrate type $A$ between two wedges of type $B$ (see Fig.\ref{Fig.7.}). Following
our earlier treatment of heterogeneous planar substrates we suppose that exactly
at saturation chemical potential $\mu \! = \! \mu_{sat}(T)^{-}$ the
filling height on the $B$ wedge is finite but that an infinite domain of $A$ 
wedge would be completely filled. This geometry induces the condensation of a
large droplet of liquid over the $A$ region with the interface height pinned to
finite values at the end points (taken to be  $y \! = \! 0$ and $y \! = \! L$). As before there is
no volume constraint on the amount of liquid adsorbed so that as 
the maximum height of the droplet $\ell_{0}^{max} \! \rightarrow \! \infty$ 
as $L \! \rightarrow \! \infty$ and we can anticipate scaling 
behaviour for the position dependence of the equilibrium profile $\langle \ell_{0}(y) \rangle$. The 
details of the partial filling of the $B$ region are not important for the 
scaling structure of the droplet shape and there are a number of specific ways
of inducing the pinning. For example if the contact angle of the liquid drop on 
the planar $B$ substrate $\theta_{B} \! > \! \pi/2$ then the associated wedge is never filled (by liquid) for
any temperature. If $\theta_{B} \! < \! \pi/2$ then one has to chose substrate $A$ such that 
$\theta_{A} \! < \! \theta_{B}$ so that the $A$ wedge fills before the $B$ wedge. 
We now distinguish between two distinct manifestations of scaling for the finite-size droplet shape:

{\bf{(A)}} Droplet shapes exactly at the filling transition temperature.

For this case the temperature $T$ is tuned exactly to the filling transition
of the fluid on the (infinite) $A$ wedge, i.e. the temperature satisfies $\theta_{A}(T_{F}) \! = \! \alpha$. 
This is the marginal value of the temperature at which the maximum
droplet height $\ell_{0}^{max}$ becomes macroscopic as $L \! \rightarrow \! \infty$.
 
{\bf{(B)}} Droplet shapes above the filling transition temperature.

If $T \! > \! T_{F}$ then the droplet height also grow to infinity as $L \! \rightarrow \! \infty$. However
because of the relevance of the linear term in the filling binding potential (\ref{54}) we cannot assume 
that the scaling of the droplet shape is the same as that obtained exactly at $T \! = \! T_{F}$. 
Indeed droplet shapes for $T \! > \! T_{F}$ may be viewed as finite-size effects occurring at the complete-filling 
transition. However fluctuations here do not have the same quasi-one-dimensional characteristics of critical filling. 
Analysis of the critical exponents at complete filling show that $\xi_{x}$ and $\xi_{y}$ diverge with the same critical 
exponent as $\mu \! \rightarrow \! \mu_{sat}^{-}(T)$ \cite{9,10}. Consequently the 1-$d$ model (\ref{53}) is not justified 
for this transition and analysis of the droplet shape in the present heterogeneous wedge problem must be based on the full 
model (\ref{51}). This is a much more difficult problem and we will not consider this here.

Our treatment of droplet shapes in the wedge parallels that of droplet shapes on planar substrates. The 
effective Hamiltonian on the heterogeneous wedge is taken to be the generalisation of the filling model (\ref{53})
\begin{equation} \label{59}
H_{F}[\ell] = \int dy \left \{ \frac{\Sigma \ell_{0}}{\alpha} \left (\frac{d \ell_{0}}{d y} \right)^{2} + V(\ell; y) \right \}
\end{equation}
where $V(\ell; y)$ is a position dependent direct filling potential that changes from being of type $V_{A}(\ell)$ 
to $V_{B}(\ell)$ as one crosses the heterogeneous region. The Ward identity for this model is again easily derived. We find
\begin{equation} \label{60}
\frac{2 \Sigma}{\alpha} \, \langle \ell_{0}(y) \, \ell_{0}^{\prime \prime}(y) \rangle = \left \langle \frac{\partial V^{\prime}}{\partial \ell}(\ell; y) \right \rangle + \frac{\Sigma}{\alpha} \, \left \langle \ell_{0}^{\prime}(y)^{2} \right \rangle
\end{equation}
so that within the droplet we can write
\begin{equation} \label{61}
\frac{2 \Sigma}{\alpha} \, \langle \ell_{0}(y) \, \ell_{0}^{\prime \prime}(y) \rangle - \frac{\Sigma}{\alpha} \, \langle \ell_{0}^{\prime}(y)^{2} \rangle = \langle V_{A}^{\prime}(\ell) \rangle. 
\end{equation}
To evaluate the RHS we again resort to the concept of an effective potential which we showed to yield exact results for the 
wetting droplet in $1 + 1$ dimensions. Thus we seek to solve the field equation (writing $\langle \ell_{0} \rangle \equiv \ell_{0}$, for convenience)
\begin{equation} \label{62}
\frac{2 \Sigma}{\alpha} \, \ell_{0} \, \ell_{0}^{\prime \prime} + \frac{\Sigma}{\alpha} \, \ell_{0}^{\prime 2} = V_{ef\!f}^{\prime}(\ell_{0})
\end{equation} 
using the effective potential $V_{ef\!f}$ evaluated at $T \! = \! T_{F}$ for short-ranged and 
long-ranged forces. For long-ranged forces we will assume that $p \! = \! 2$ as is appropriate for realistic solid-fluid 
interfaces. This differential equation is easily integrated using standard techniques and we only quote the final results.

For droplet scaling exactly at $T_{F}$ and short-ranged forces we find
\begin{equation} \label{63}
\langle \ell_{0}(y) \rangle = \ell_{0}^{max} 4^{1/3} \left [\frac{y}{L}\left (1 -  \frac{y}{L}\right) \right ]^{1/3}
\end{equation}
where $\ell_{0}^{max} \! \propto \! L^{1/3}$ denotes the mid-drop height. Note that this scaling form is precisely in accord with the 
prediction of conformal invariance for one dimensional interfaces with wandering exponent $\zeta \! = \! 1/3$. For long-ranged 
forces on the other hand integration of (\ref{62}) yields a quite different form of the scaling function. We find that 
$\langle \ell_{0}(y) \rangle /\ell_{0}^{max} \! \equiv \! \tilde{\ell}_{0}(y)$ satisfies the cubic equation
\begin{equation} \label{64}
3 \, \tilde{\ell}_{0}^{2} + \tilde{\ell}_{0}^{3} = 16 \, \frac{y}{L} \left( 1 - \frac{y}{L} \right)
\end{equation}
with $\ell_{0}^{max} \! \sim \! L^{1/2}$. Thus the scaling of droplet shapes at wedge wetting is only consistent with conformal 
invariance for systems with short-ranged forces.

\section{Conclusion}
In this paper we have investigated the scaling behaviour of finite-size
droplet shapes on heterogeneous planar substrates and wedges. Our predictions are
restricted to systems at two-phase bulk liquid-vapour coexistence (chemical
potential $\mu \! = \! \mu_{sat}$) in the grand canonical ensemble, i.e. without volume constraint. The 
main conclusions of article are the following:

{\bf{(A)}} For three-dimensional systems with short-ranged forces the scaling of the droplet shape and its height-height correlations 
can be understood using conformal invariance. In fact using standard wetting renormalization group ideas the conformal invariance may 
be explicitly proved for such systems. Consequently the study of droplet shapes in laboratory systems with effective short-ranged 
forces such as polymers, superconductors, liquid metals and even some fluid-fluid interfaces may be viewed as a possible experimental 
testing ground for many of the ideas developed in conformal-field theory as well of course as the fluctuation theory of fluid interfaces.

{\bf{(B)}} For three-dimensional systems with long-ranged (van der Waals) dispersion forces the droplet shape still exhibits a strong 
degree of scaling invariance related behaviour even though there is no longer any conformal invariance.

{\bf{(C)}} For model systems of interfaces in two-dimensional bulk geometries with short-ranged forces the scaling shape of the droplet 
on the heterogeneous wall can be understood quantitatively using an effective potential accounting for the entropic repulsion of the 
interface from the wall and is completely in accord with the predictions of conformal invariance for one dimensional field theories.

{\bf{(D)}} Finite-size droplets in wedge geometries show a number of interesting scaling behaviours depending on whether the 
interfacial pinning is done exactly at or above the filling transition temperature. Exactly at $T_{F}$ and for systems with 
short-ranged forces the prediction of conformal invariance for the shape function is again valid. For long-ranged forces a 
different scaling behaviour is observed.

The predictions of this paper are based entirely on the analysis of effective interfacial models. It would be interesting to test 
these ideas using more microscopic approaches such as density-functional theory. The density (or magnetisation) profile for such 
systems is of course highly inhomogeneous and even analysis of a Landau-like model would require numerical methods. Beyond mean-field 
Ising model simulations could test the role of the wetting parameter in determining the scale of the droplet height. Finally we hope 
that at least some of our predictions are open to experimental verification from observation of droplet shapes in the micro-metre range.

E.D.Macdonald acknowledges the EPRSC for a studentship and C.R. acknowledges the financial assistance of the E.C. under contract ERBFMBICT983229.

\begin{appendix}
\section{}
We show here that, in two dimensions, the equation
\begin{equation}
\label{A1}
\nabla^2\ell=C\,e^{-\tilde{\kappa}\,\ell},
\end{equation}
(where $C$ and $\tilde\kappa$ are real numbers) 
is conformally invariant. In particular, we
demonstrate that Eq.\ (\ref{A1}) is invariant
under the following conformal transformation
\begin{eqnarray}
\nonumber
(x,y) & \longrightarrow & (u,v)\\
e^{-\tilde{\kappa}\,\ell(x,y)}& \longrightarrow &
\label{A2}
e^{-\tilde{\kappa}\,\overline{\ell}(u,v)}\,|w'|^{\,x_b}
\end{eqnarray}
where $(u,v)$ are given by the relation (in the
complex plane)
$u+iv=w(z)$, $z\equiv x+iy$, $w(z)$ is
any analytic function, $|w'|$ is the local length rescaling factor 
of the transformation and $x_{b}=2$.

If we define
\begin{eqnarray*}
\nabla\equiv\left(\frac{\partial}{\partial x},\frac{\partial}{\partial y}\right)
\hspace{2cm} \\
\overline{\nabla}\equiv\left(\frac{\partial}{\partial u},\frac{\partial}{\partial v}\right)
\hspace{2cm} \\
J\equiv|w'(z)|^{2}=\left(\frac{\partial u}{\partial x}\right)^{2}+
\left(\frac{\partial v}{\partial x}\right)^{2},
\end{eqnarray*}
substitution of (\ref{A2}) into (\ref{A1}) yields
\begin{eqnarray}
\label{A3}
\nabla^2\overline{\ell}-\frac{x_{b}}{2\tilde{\kappa}}\;\nabla^2\log J =
C\;e^{-\tilde{\kappa}\,\overline{\ell}(u,v)}\;J^{\,x_{b}/2}.
\end{eqnarray}
Using analytic properties of the conformal transformation
\begin{eqnarray*}
\frac{\partial u}{\partial x}=\frac{\partial v}{\partial y}, \hspace{1cm}
\frac{\partial u}{\partial y}=-\frac{\partial v}{\partial x}
\end{eqnarray*}
and standard methods, it can be shown that $\nabla^2\log J = 0$ and 
$\nabla^2 = J\;\overline{\nabla}^2$. Therefore, eq.\ (\ref{A3})
transforms into
\begin{eqnarray*}
J\;\overline{\nabla}^2\,\overline{\ell} =
C\;e^{-\tilde{\kappa}\,\overline{\ell}(u,v)}\;J^{\,x_{b}/2},
\end{eqnarray*}
which, in turn, recovers (\ref{A1}) for $x_{b}=2$.
\end{appendix}

\end{document}